\documentclass[aps,twocolumn,pre,superscriptaddress,notitlepage]{revtex4-2}

\usepackage[utf8]{inputenc}

\usepackage{epsfig}
\usepackage{natbib}
\usepackage{xcolor}
\usepackage{amssymb,amsthm,amsmath}
\usepackage{wasysym}

\newcommand{\br}{{\bf r}}
\usepackage{hyperref}
\usepackage[normalem]{ulem}

\newcommand{\SM}{\textit{Supplementary Materials}}

\begin{document}

\title{Morphological routes to extinction: A mechanistic assessment of habitat loss}

\author{E. H.  Colombo}\email[Corresponding author: ]{e.colombo@hzdr.de}
\affiliation{Center for Advanced Systems Understanding -- Helmholtz-Zentrum Dresden-Rossendorf, Untermarkt 20, 02826 G\"{o}rlitz, Germany\looseness=-1}

\author{L. Menon}
\affiliation{Statistical and Interdisciplinary Physics Division, Centro Atómico Bariloche (CNEA), R8402AGP Bariloche, Argentina }
\affiliation{Department of Physics, PUC-Rio, Rua Marqu\^es de S\~ao Vicente  225, 22451-900 G\'avea, Rio de Janeiro, Brazil\looseness=-1}

\author{E. Hern\'andez-Garc\'{\i}a}
\affiliation{Instituto de F\'{\i}sica Interdisciplinar y
Sistemas Complejos (IFISC), CSIC-UIB, Campus Universitat Illes
Balears, 07122, Palma de Mallorca, Spain\looseness=-1}

\author{C. Anteneodo}
\affiliation{Department of Physics, PUC-Rio, Rua Marqu\^es de S\~ao Vicente  225, 22451-900 G\'avea, Rio de Janeiro, Brazil\looseness=-1}

\begin{abstract}
Habitat loss driven by climate and anthropogenic pressures alters patch morphology, with critical consequences for population persistence. Geometric and mechanistic metrics are commonly used to quantify degradation, yet their respective limitations remain poorly understood. Here, we address this gap using a
reaction–diffusion framework for population growth and dispersal
in a viable patch embedded in a hostile environment. 
We compare geometric descriptors of patch shape with a mechanistic metric derived from population growth near the extinction threshold. Along degradation trajectories, we find that geometric metrics systematically overestimate persistence, suggesting moderate and decelerating impacts, whereas mechanistic indicators  reveal rapid, accelerating approaches to extinction. These results highlight fundamental limitations of geometric approaches and underscore the need for mechanistic assessments when evaluating biodiversity loss in complex landscapes.
\end{abstract}

\maketitle

\section{Introduction}
Habitat degradation, often caused by human activities or environmental changes, is considered a major driver of biodiversity loss~\citep{Fahrig2003, Haddad2015}. 
A central component of degradation is the geometric transformation of habitats: reductions in total area, increases in perimeter-to-area ratios, and changes in patch shape and connectivity alter the spatial structure of populations and the ecological processes they support~\cite{turner2001landscape}. 
In recent decades, the combined pressures of climate change, land-use conversion, and other anthropogenic stresses have produced smaller and more irregular habitat patches, amplifying edge effects, reducing connectivity, and increasing species extinction risk~\cite{zou2025}. Consequently, characterizing degradation and assessing its consequences for species persistence remain pressing challenges.

Landscape alterations have most often been quantified using geometric metrics that capture the size, shape, and configuration of habitat patches~\cite{turner2001landscape}. Total area is the simplest and most widely used metric, reflecting resource availability. Area-to-perimeter ratios go a step further, capturing the trade-off between habitat area and exposure to boundary effects, such as increased predation, edge microclimate, and human disturbance~\cite{Murcia1995, Woodman2025}.

These descriptors effectively summarize landscape structure and its transformation over space and time, and have guided conservation planning, identified fragmentation hotspots, and informed land-use policies~\cite{Pullin_2002}. However, purely geometric metrics do not directly translate into a mechanistic understanding of how population dynamics respond to habitat morphology~\cite{mcgarigal2006landscape,wu2004effects,fahrig2017ecological}, nor do they capture the balance between local population growth and boundary-mediated losses that governs persistence~\cite{hanski2000metapopulation}. As a result, forecasts of species persistence may be misleading when based solely on such metrics~\cite{li2004use,fahrig2017ecological}.

Theoretical advances have established mechanistic links between landscape structure and species persistence. For a single patch surrounded by hostile conditions, the dominant eigenvalue of the reaction–diffusion operator governing population growth and dispersal serves as an indicator of viability, determining whether demographic gains can offset dispersal losses and boundary-driven mortality at low densities~\citep{skellam1951random, holmes1994partial, CantrellCosner2003}. In heterogeneous multi-patch landscapes, eigenvalue-based metrics such as metapopulation capacity~\cite{hanski2000metapopulation} have been widely used to quantify extinction thresholds under habitat loss and fragmentation~\citep{zou2025}. Together, these approaches provide a distinct perspective on persistence by integrating landscape configuration with population dynamics, thereby capturing emergent (functional) connectivity~\cite{calabrese2004comparison}.

Despite widespread use, the relationship between eigenvalue-based metrics and traditional geometric descriptors remains only partially understood. This gap is critical, as previous studies have shown that habitat amount and simple geometric indices can fail to predict persistence when compared with mechanistic measures such as metapopulation capacity or reaction–diffusion eigenvalues~\citep{CantrellCosner2003, wu2004effects,perry2005experimental,mcgarigal2006landscape,fahrig2017ecological}. In particular, landscapes with identical area or similar perimeter–area relationships can exhibit markedly different persistence outcomes due to differences in spatial configuration and boundary effects. However, these discrepancies have mostly been identified in case-specific or empirical settings, lacking a controlled framework for systematically comparing geometric and mechanistic metrics across a well-defined space of habitat morphologies. As a result, when and to what extent geometric measures misrepresent degradation remains unresolved.


In this work, we clarify discrepancies between geometric- and mechanistic-based predictions of population persistence
across realistic patch configurations with varying area and boundary roughness. Within the morphological space defined by these two properties, we introduce degradation paths, along which mechanistically informed metrics indicate 
stronger trends toward extinction than geometric metrics. 
Although real population dynamics are more complex than the minimal model considered here, placing  geometric and mechanistic approaches side by side reveals clear and systematic differences between them. These differences are critical when designing land-use regulations and management strategies~\cite{banks2020countering}.

\section{Methods}
\label{sec:method}

\subsection{Population dynamics}
\label{sec:population}

We consider a population of individuals that move randomly, reproduce, and die~\cite{skellam1951random}. The space they can explore is a  two-dimensional  continuous patch of landscape where conditions for growth are met~\cite{turner2001landscape}. At the boundary, however, individuals find hostile conditions that, as a simplifying hypothesis, we assume lead to immediate death.  
Under this setting, the scalar field  $\rho(\br,t)$, which describes the population density at any location  $\br=(x,y)$  and time $t$, can be modeled by a reaction-diffusion process comprising diffusion and growth terms.

\begin{figure}[ht!]
\centering
\includegraphics[width=1\columnwidth]{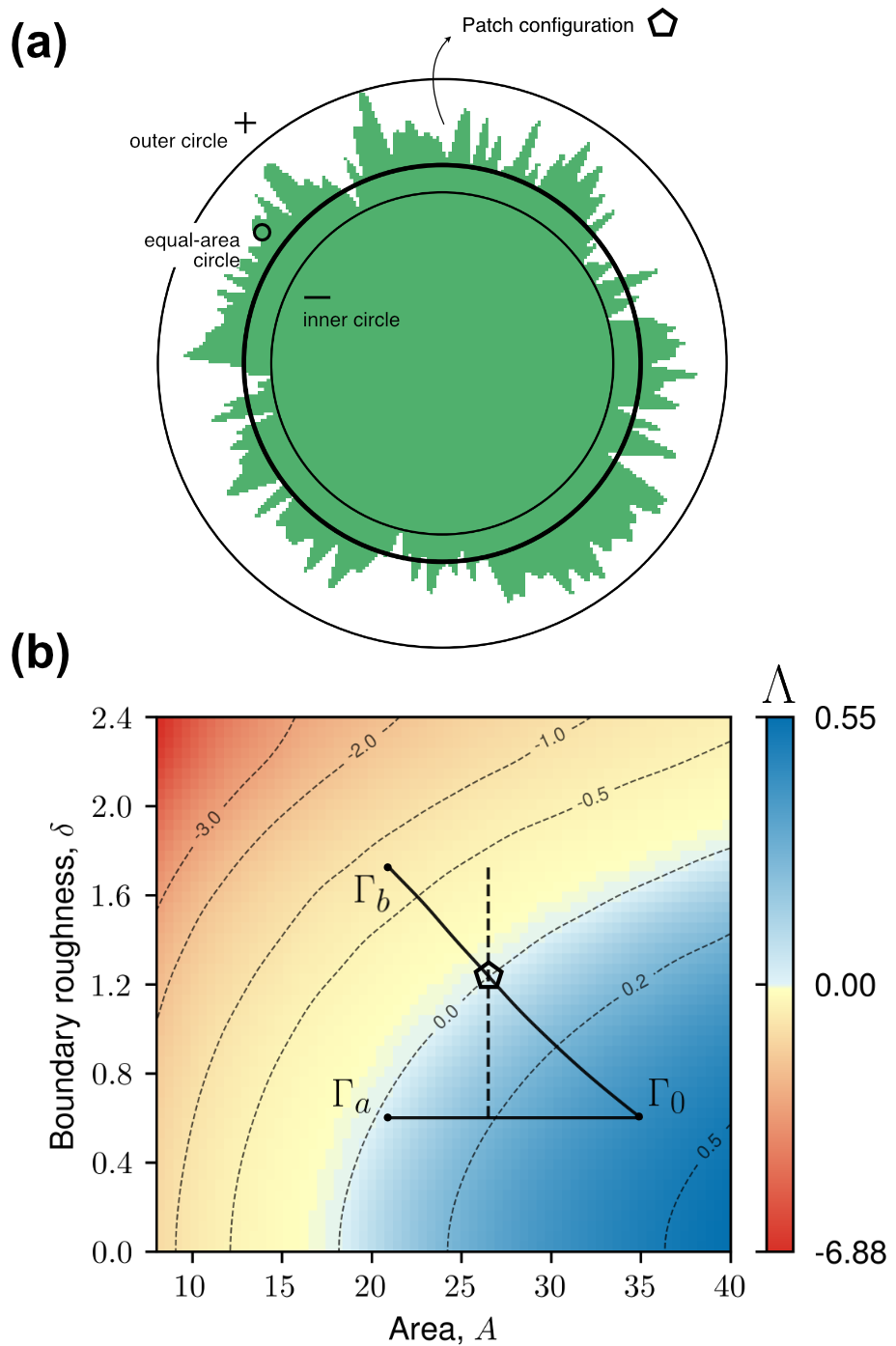}
    \caption{\textbf{Patch morphological space and degradation routes.}
(a) Habitat patch constructed using Eq.~(\ref{eq:fractal_boundary}) with intermediate boundary roughness, $(A,\delta)\simeq(26.5,1.2)$, with its deviations from the circular patch with the same area (o) bounded by the inner ($-$) and outer ($+$) circles. b) Patch morphological space and degradation paths. Heatmap  and labeled contours indicate  the values of the dominant eigenvalue $\Lambda$ (Eq.~\ref{eq:lambda}) across a range of values of patch area, $A$, and boundary roughness, $\delta$. Reference configuration $\Gamma_0=(35, 0.6)$ and focal configurations $\Gamma_a=(21, 0.6)$ (reached from  $\Gamma_0$ by decreasing area without boundary transformation) and $\Gamma_b=(21, 1.8)$ (reached via maximum boundary transformations) are highlighted, as well as the path connecting them. Contour lines at different values of $\Lambda$ are indicated by thin dashed lines, with $\Lambda=0$ indicating the extinction threshold. Intermediate configuration on the extinction threshold of patch $\Gamma_0\to \Gamma_b$ ($\pentagon$) is reproduced in panel a). Vertical path with fixed area is shown with dashed thick black line crossing the $\pentagon$ configuration. The model parameters  in Eq.~(\ref{eq:main}) 
take the values 
are $a=D=1$.
} 
    \label{fig:main}
\end{figure}

The reaction-diffusion equation can be complicated, incorporating density-dependent, nonlocal, and nonlinear effects, e.g. 
$\partial_t \rho = \nabla(D(\rho)\nabla\rho) + f(\rho)$. Nevertheless, 
when investigating the stability of the extinction state ($\rho\to 0^+$), the crucial component needed is the linearized version of the dynamics. Therefore, regardless of the specific mechanisms underlying birth-death and movement processes, in the near-extinction limit, $\rho \approx 0$, with few exceptions~\cite{COLOMBO201811},  we expect the linearized dynamics given by 
\begin{align}\notag
    &\,\partial_t \rho = \mathcal{L}[\rho] + \mathcal{O}(\rho^2) \,,  \ \mathcal{L} \equiv  D\nabla^2 + a ,\\[0.5em] 
    &\textrm{with}\, \rho(\mathbf{r}\in \mathcal{B}, t) = 0
\label{eq:main}
\end{align}
to provide a correct identification of the extinction threshold. $D$ is a linear diffusion coefficient and $a$ is a linear growth rate. The boundary $\mathcal{B}$ separates the habitat patch from the surrounding matrix (respectively green and white regions in Fig.~\ref{fig:main}a). Consequently, locations along $\mathcal{B}$ are exposed to unfavorable environmental conditions that abruptly reduce population persistence. In Eq.~(\ref{eq:main}), we approximate this effect by assuming that individuals die immediately upon crossing the boundary. Mathematically, this is implemented through a Dirichlet boundary condition imposing zero population density along $\mathcal{B}$.
In our synthetic experiments, we will employ a fractal-like patch model (Sec.~\ref{sec:patchmodel}) to generate $\mathcal{B}$ under different parameter settings controlling both patch area and boundary roughness.

\subsection{Patch morphology}
\label{sec:patchmodel}

The scenarios we consider assume 
a complex-boundary model to generate distinct patch boundaries,  $\mathcal{B}$  (see example in Fig.~\ref{fig:main}a). Importantly, the patch model allows for the control of key morphological features. Specifically, morphological configurations  are characterized by a point $\Gamma$ in a morphological two-dimensional space 
$\Gamma=(A,\delta)$, where $A$ is the patch area and $\delta$ characterizes the boundary roughness.

 We parameterize the boundary by using polar coordinates centered at a point in its interior, with $b$ the radial coordinate, and $\theta$ the angular coordinate. Thus, the boundary $\cal B$, defined in Eq.~(\ref{eq:main}), is given by the closed curve defined by the function $b(\theta)$, for $\theta\in (0, 2\pi)$.  Explicitly, for each morphological configuration $\Gamma = (A,\delta)$ we define our fractal-like boundary choosing
\begin{align}\notag
    &\mathcal{B}(A,\delta) = \{ b(\theta| A,\delta)\, ;\; \forall \theta \in [0,2\pi] \}\quad \textrm{with,}\\
    &b(\theta| A,\delta) = \sqrt{\frac{A}{\bar{A}(\delta)}}\, \left[1 + \delta\zeta(\theta)\right],
    \label{eq:fractal_boundary}
\end{align}
where $\zeta$ is a pink noise  defined below  and $\bar{A}(\delta) = (1/2)\int_0^{2\pi} [1+\delta\zeta(\theta')]^2d\theta'$  is a normalization factor ensuring that the area is always $A$, regardless of the specific shape of $\zeta$.
 
The choice of $\zeta$ is inspired by natural boundaries, such as coastlines~\cite{mandelbrot1967long}, islands, corals~\cite{bradbury1982fractal,gimenez2025unravelling}, and forest patches~\cite{PhysRevE.109.L042402}, which exhibit fractal-like features. In such cases, empirical evidence suggests a  Fourier-spectrum power-law structure, which we model as
\begin{equation}
\zeta(\theta) =    
\sum_{n=n_0}^{n_M} a_n \sin(n\theta + \phi_n)\Big/\sum_{n=n_0}^{n_M} a_n ,
\end{equation}
with $a_n = 1/\sqrt{n}$ denoting the  Fourier amplitudes and 
$\phi_n \in [0,\pi]$ random phases. 
We set $n_0=2$ and $n_M=100$. 

\subsection{Degradation indices}
\label{sec:degradation}
To quantify the reduction of population persistence capability on a patch, we compare a focal morphological configuration, $\Gamma = (A,\delta)$, to a reference one, $\Gamma_0 = (A_0,\delta_0)$. For a given geometric- or mechanistic-based metric $H$, that characterizes population persistence capability, e.g. the patch area, we define a degradation index  $\mathcal{D}_H$ as

\begin{equation}
    \mathcal{D}_H(\Gamma|\Gamma_0) = \frac{H(\Gamma_0) - H(\Gamma)}{H(\Gamma_0)} \times 100\, .
    \label{eq:degradation}
\end{equation}
This structure allows for straightforward interpretation of how shape transformations affect population persistence. In particular, the value of $\mathcal{D}_H$ reflects the percentage of $H$ lost as the patch transitions from the reference configuration $\Gamma_0$ to the focal one, $\Gamma$. For example,  $\mathcal{D}_H=50\%$  indicates that half of the original persistence capability was lost. Negative values of $\mathcal{D}_H$ indicate a gain in persistence capability. 
Below, we define metrics for the geometric and mechanistic metrics that can be plugged into Eq.~(\ref{eq:degradation}) to quantify the loss of persistence associated with patch morphological transformations.

\subsubsection{Geometric-based approach}

Geometric indices of degradation are largely constructed from measures
 of patch area and boundary properties such as its perimeter~\cite{turner2001landscape}. The combination of both of them given by the area-to-perimeter ratio, 
\begin{equation}
    H_{\textrm{AP}} = \frac{A}{P}\, ,
    \label{eq:asp1}
\end{equation} 
where $P$ is the patch perimeter of the boundary $\cal B$,  has a special advantage. Despite its simplicity, $A/P$ captures both the positive effect of patch area and the negative effect of edge exposure — the two fundamental mechanisms behind extinction risk. Consequently, it has been widely used across different ecological contexts~\cite{hargis1998behavior,helzer1999relative}.
Variations of the area-to-perimeter ratio, accounting for different trade-offs between area and boundary effects and shape templates, have also been proposed~\cite{bogaert2000alternative}. For completeness, we also include the cases 
\begin{equation}
    H_{\textrm{AP2}} = A/P^2\quad \textrm{and }\quad H_{\textrm{2AP}} = \sqrt{A}/P\, .
    \label{eq:ap2}
\end{equation}

\subsubsection{Mechanistic-based approach}

In order to mechanistically assess the trade-off between the outcomes produced by the population dynamics in the bulk and at the boundary of the patch, we numerically solve the  linear  reaction-diffusion problem described in Eq.~(\ref{eq:main}). More specifically, we focus on the dynamics near extinction, $\rho \approx 0$. Under this limit, the linear dynamics in Eq.~(\ref{eq:main}) describes a much broader class of models.  The solution of (\ref{eq:main}) can be written as an eigenfunction expansion (see \SM{}), 
\begin{align}
\label{eq:expansion}
 \rho(\mathbf{r},t) &= \sum_{i}^{\infty} S_i(\mathbf{r})e^{ \lambda_i t} \sim S_0(\mathbf{r})e^{\Lambda t}  \, ,
\end{align}
which  has been approximated at long times  by the term associated with the dominant eigenvalue $\Lambda= \max(\{\lambda_i\})$ of the operator $\cal L$ defined in Eq.~(\ref{eq:main}).

Population persistence capability can then be characterized directly via the maximum eigenvalue, 
\begin{equation}
\label{eq:lambda}
    H_{\textrm{ME}} = \Lambda  .
\end{equation}

 If $\Lambda< 0$, then all exponential components in Eq.~ (\ref{eq:expansion}) decay with time, and the population goes extinct at long times. On the other hand, if $\Lambda>0$, then at least one rate is positive, and the population tends to grow if placed near extinction. Therefore, the critical point $\Lambda=0$ sets the extinction threshold.

Exact calculations  for simple shapes, such as  circular or square patches, have been derived in early works~\cite{skellam1951random}, providing first insights into how eigenvalues change as a function of the area and boundary of the patch.  For these two-dimensional shapes, extinction occurs when the area is below a critical patch area, $A_c$, which from dimensional analysis should depend on growth rate $a$ and diffusion coefficient $D$ as  $A_c = g_c~D/a$. When using dimensionless units, the expression reduces to $A_c=g_c$ so that $g_c$ is a factor that depends only on the geometrical properties of the patch, and not on the dynamical properties $a$ and $D$. $g_c$ minimum for a circle and increases as we deviate from circularity at constant area~\cite{holmes1994partial}. Consequently, population growth in circular patches is always larger than in any other shape of the same area, and would persist until lower area values.  
In general, it has been shown that, for simply connected patches (without internal voids) with arbitrary boundaries, the dominant eigenvalue $\Lambda$ is bounded by those corresponding to the inscribed ($-$) and circumscribed ($+$) circles, $\Lambda \in [\Lambda_-, \Lambda_+]$~\cite{bookcantrell2004spatial}, which define the ring that contains the boundary (see Fig.~\ref{fig:main}a). Several particular cases have been analyzed, considering, for example, spatio-temporal  variations in patch morphology~\cite{Cantrell2001,Neicu2000,Lin2004,Ballard2004,Colombo2016}, population advection~\cite{Pachepsky2005,Ryabov2008,dornelas2024movement}, chemotaxis~\cite{kenkre2008nonlinearity}, nonlinear responses~\cite{colombo2018nonlinear}, and two-sex populations~\cite{fonseca2013modeling}. However, in all these cases, the exact values of $\Lambda$ for arbitrary shapes beyond simple geometries can only be obtained numerically~\cite{cantrell2001predator}.

In order to produce results across a range of morphological configurations, we use a  continuous-limit extrapolation method  to accurately and efficiently extract the value of $\Lambda$. The method combines standard  eigenvalue calculation  schemes with Richardson's extrapolation~\cite{richardson1911ix} to  minimize  the effect of discretization (see \SM{}).

For a given morphological configuration $\Gamma$, population growth rate $a$ and diffusion constant $D$, 
the corresponding largest eigenvalue can be denoted by 
$\Lambda(\Gamma; a, D)$. 
Without loss of generality, eigenvalue numerical calculations will be performed in the following by fixing  $a=1$ and $D=1$ (in arbitrary units), so that unless otherwise stated, the notation $\Lambda$ will in fact denote the eigenvalue $\Lambda(\Gamma; 1,1)$. Results for other values of $a$ and $D$ can then be obtained via the scaling relation
\[
\Lambda(\Gamma; a, D) = a + D\big[\Lambda(\Gamma; 1,1) - 1\big].
\]
The implications of this transformation for the degradation index (Eq.~\ref{eq:degradation}) can be directly propagated, yielding
\[
\mathcal{D}_{H_\mathrm{ME}}(\Gamma|\Gamma_0; a, D)
=
\frac{\mathcal{D}_{H_\mathrm{ME}}(\Gamma|\Gamma_0; 1,1)}
{1 + \left(\frac{a}{D} - 1\right)\Lambda(\Gamma_0;1,1)^{-1}}.
\]

\section{Results and Discussions}

\subsection{Morphological space}

To first assess the effects of different morphological configurations on population persistence, we compute the mechanistic persistence capability $H_{\textrm{ME}}=\Lambda$, Eq.~(\ref{eq:lambda}), across a range of morphological configurations (Fig.~\ref{fig:main}b). 
We treat this as the ground truth of the resulting dynamics, given the assumptions and limitations of the minimalist model defined in Eq.~(\ref{eq:main}).

Exploration of the morphological space reveals a clear curved boundary separating the persistence and extinction phases (contour line for $\Lambda = 0$ in Fig.~\ref{fig:main}b). The shape of this boundary indicates that moving diagonally toward lower area and higher boundary roughness (i.e.,  simultaneously leftward  and upward in the heatmap) leads to the  fastest decrease in $\Lambda$, reflecting the combined effects of area loss and boundary-mediated mortality.

\subsection{Morphological paths}
\label{sec:paths}

Due to the presence of strong gradients of $\Lambda$ in morphological space, paths defined by sequences of shape transformations can have distinct consequences.
First, extinction may occur when moving leftward in the heatmap (Fig.~\ref{fig:main}b), as a critical minimum area is crossed.
This corresponds to the simpler case of area loss with boundary roughness held constant.
More interestingly, increasing boundary roughness (dashed line moving upward in Fig.~\ref{fig:main}b) can, by itself, induce a transition from persistence to extinction.

\begin{figure}[b!]
    \centering
\includegraphics[width=0.5\textwidth]{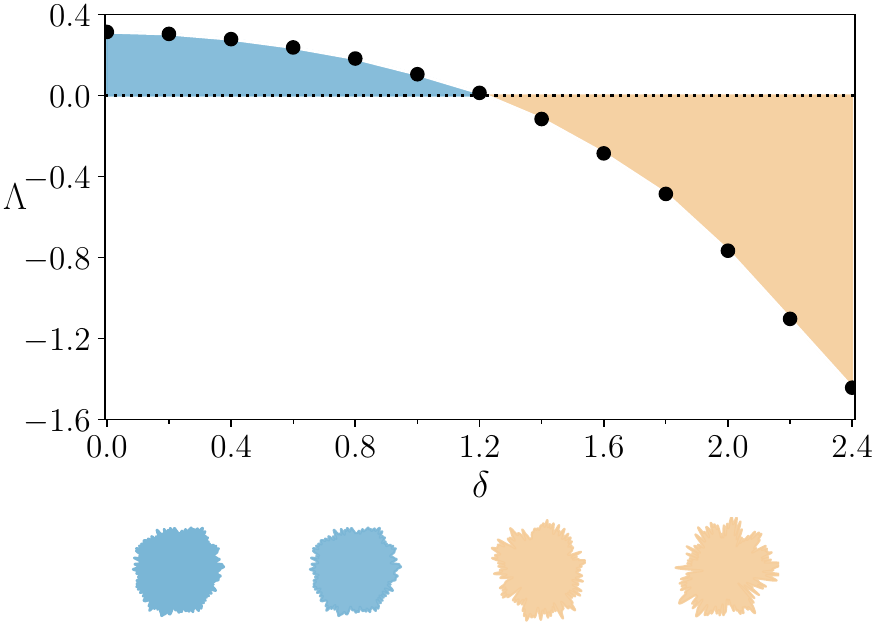}
    \caption{ {\bf How boundary roughness impacts on persistence.} Dominant growth rate $\Lambda$ as a function of boundary roughness, $\delta$, for area $A=26.5$ (along the dashed vertical line in Fig.~\ref{fig:main}b). Patch shapes, as defined by Eq.~(\ref{eq:fractal_boundary}), with boundary roughness $\delta=$  0.2, 0.8, 1.4 and 2.0 are depicted. 
}
    \label{fig:delta}
\end{figure}

To contrast predictions from the mechanistic framework with geometric ones, we defined paths in morphological space. Neglecting boundary transformations, we considered a path connecting configurations $\Gamma_0$ to $\Gamma_a$, along which the boundary roughness $\delta$ remained fixed while the area varied. For comparison, we also generated a path connecting $\Gamma_0$ to $\Gamma_b$ using a  steepest-descent procedure, in which the configuration was updated through small increments in $(A, \delta)$ space  in the direction minimizing  the next value of $\Lambda$, thereby tracing a path toward $\Gamma_b$ (see Fig.~\ref{fig:main}b caption for details).  Therefore, while $\Gamma_0\to \Gamma_a$ works as a null reference for possible degradation progress without boundary effects, boundary effects on persistence are maximized along the $\Gamma_0\to \Gamma_b$ path.  

These two scenarios can then serve as stress tests for the persistence metrics. Specifically, we are interested in how the additional “length” traveled, due to concomitant boundary roughening while area is lost, promotes a deficit in persistence, and, importantly, how this deficit is perceived by the geometric- and mechanistic-based metrics.

In order to parameterize the progression of degradation along the paths  $(\Gamma_0\to \Gamma_a)$ and $(\Gamma_0\to \Gamma_b)$, we define the  percentage of lost area, $100\times[A(\Gamma)-A(\Gamma_0)]/A(\Gamma_0)$, for a given configuration $\Gamma$ (belonging to one of the paths) relative to the reference configuration $\Gamma_0$. 
As the empirical values of $a$ and $D$ used to tune the mechanistic-based predictions are hardly known with precision, and the scaling factors behind geometric-based results can only be heuristically estimated,  we use the degradation indices of Sec.~\ref{sec:degradation}, which characterize  degradation in terms of percentages, as an appropriate way to obtain results sensible only to the essential characteristics of the domain shapes.  Along the morphological paths we calculated degradation indices in Eq.~(\ref{eq:degradation}) to measure the relative change in persistence capability (either via geometric- or mechanistic-based approaches) with respect to the reference morphological configuration $\Gamma_0$. This way we can focus on the metrics' response to shape changes rather than their  specific  value at a given instance.

\noindent
\subsection{ Limitations of geometric-based metrics}  

The first clear difference between the two approaches, regardless of the path considered, lies in the range of the metrics. While the mechanistic metric is unbounded, $\Lambda \in (-\infty, \infty)$, geometric metrics are strictly positive, e.g., $A/P > 0$. This difference highlights a key advantage of the mechanistic approach: it allows the extinction threshold to be defined explicitly once the population dynamics—through parameters such as $a$ and $D$—are specified. In contrast, for geometric metrics, the definition of an extinction threshold is inherently arbitrary.
 
\begin{figure}
    \centering
\includegraphics[width=1\columnwidth]{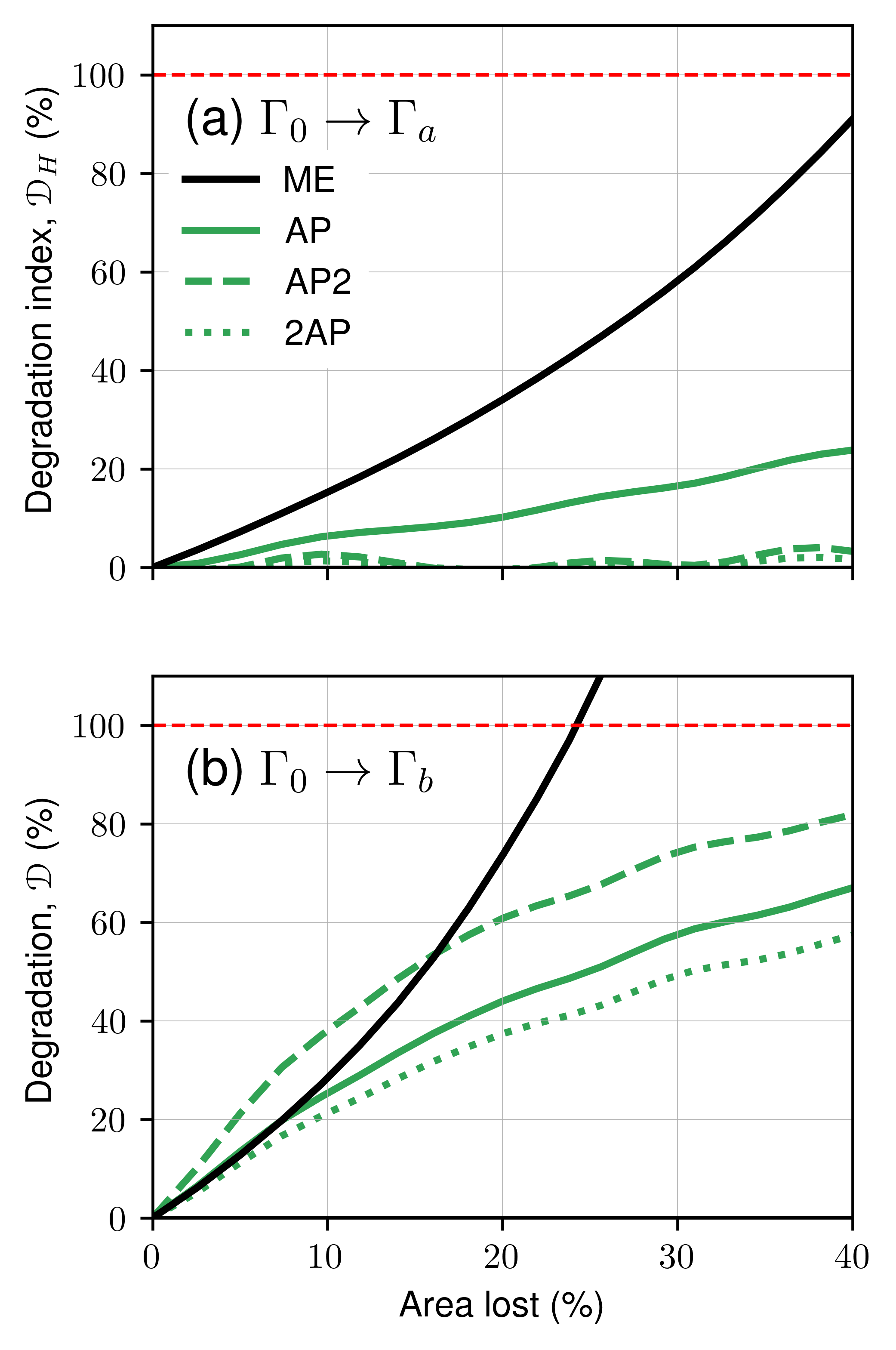}
    \caption{\textbf{Degradation along morphological paths.} Paths are parameterized by the area lost with respect to the reference configuration $\Gamma_0$ in Fig.~\ref{fig:main}. 
    a) Degradation indices vs. area lost along fixed-roughness area loss path ($\Gamma_0\to \Gamma_a$) and maximum degradation path ($\Gamma_0\to \Gamma_b$, see Sec.~\ref{sec:paths} for details). Degradation indices were calculated from Eq.~(\ref{eq:degradation}), using the mechanistic metric $\Lambda$ in Eq.~(\ref{eq:lambda}) and different geometric ones based on the area-to-perimeter ratio in  Eqs.~(\ref{eq:asp1}) and (\ref{eq:ap2}).} 
    \label{fig:degradation}
\end{figure}

Deeper contrasts between the approaches are revealed when considering degradation along the specific paths analyzed here. For fixed boundary roughness ($\Gamma_0 \to \Gamma_a$ in Fig.~\ref{fig:main}b), geometric metrics substantially underestimate the impact of degradation (green lines in Fig.~\ref{fig:degradation}a).  In fact, degradation along $\Gamma_0 \to \Gamma_a$ can become constant
(e.g., if $A \propto P^2$ for $H_{\textrm{AP2}}$ or $H_{\textrm{2AP}}$, as for our case, but also true if $A \propto P$ and $H_{\textrm{AP}}$ is selected)~\cite{bogaert2000alternative}. In any case, the existence of such a strong contrast between both approaches to degradation indices highlight that the tug-of-war between area and boundary effects can not be captured by simple geometrical relations.
Besides strong quantitative underestimation by the geometric-based degradation indices, it also misses that degradation along the boundary-preserving path ($\Gamma_0 \to \Gamma_a$) is ``accelerating''.

Following the fastest path to extinction,  {$\Gamma_0\to \Gamma_b$}, further differences between the two classes of metrics emerge, as this path involves simultaneous changes in both area and boundary properties. Along  {$\Gamma_0\to \Gamma_b$}, geometric metrics consistently indicate slow, decelerating degradation, whereas the mechanistic metric reveals fast and accelerating degradation (Fig.~\ref{fig:degradation}b). Crucially, close to $23\%$ area lost, the mechanistic-based index was capable of indicating the population extinction threshold.

\section{Final remarks}
Our synthetic experiments provide a controlled environment in which the contrasting predictions of geometric- versus mechanistic-based degradation metrics become evident. Strong qualitative differences between the two approaches are observed. Overall, the mechanistic-based degradation index indicates a more pessimistic scenario, where degradation rates are faster and accelerating.
Moreover, when mechanistic-based models are informed by empirical diffusion and growth rates from case-study populations, they can identify regions in morphological space where populations can persist (bluish region in Fig.~\ref{fig:main}b).

Although our results are obtained under simplified population dynamics and a realistic but still reduced patch model, they address key questions in population management and conservation. Furthermore, we expect the reported phenomena to be robust for a broad class of models, although exceptions may exist. Consequently, these same simplifications that helped us gain conceptual clarity also limit the applicability of our findings to the complex, multi-faceted realities of natural landscapes.

For example, landscapes are changing at a rapid pace~\cite{zou2025}, at rates that can become comparable to population dynamics. As a result, understanding population persistence from single static snapshots of the landscape may be too narrow an approach to characterize degradation~\cite{fletcher2026resurrecting}. In addition, we have neglected the fact that landscapes in many cases are composed of connected patches that create a metapopulation structure~\cite{hanski2000metapopulation}; that the presence of other species can substantially alter the stability of the overall dynamics~\cite{allesina2012stability,galla2018dynamically}; and that nonlinear diffusion and growth can reverse stability conditions in counter-intuitive ways~\cite{colombo2018nonlinear,rosa2026general}.

Regarding patch shape, although our model captures several realistic features expected in natural environments, it remains low-dimensional compared to the range of possible transformation paths observed in real landscapes. In particular, boundary spatial correlations—fixed in our case via pink noise—can vary in both space and time under natural conditions~\cite{bradbury1982fractal,sugihara1990applications}. Moreover, abrupt changes in patch geometry may arise from human activities such as deforestation~\cite{broadbent2008forest}, while urbanization processes reshape landscapes in diverse and complex ways~\cite{liu2016relationship}.

These issues can, however, be addressed by constructing customized models depending on the case of interest. Efforts to bridge theoretical and empirical approaches should indeed focus on context-specific applications where patch shape can be extracted from satellite imagery at a given resolution and population dynamics of the focal species are well documented. The methods detailed in the \SM{} can then be applied using a context-specific set of parameters. Numerical analyses could be performed on processed images to provide accurate estimates of geometric- and mechanistic-based degradation indices. Empirical studies could then assess whether the theoretically predicted differences between degradation indices, as revealed in our synthetic experiments, manifest in a significant manner in natural conditions.

\section*{Acknowledgments}

L.M. and C.A. acknowledge partial financial support by the 
Coordena\c c\~ao de Aperfei\c coamento de Pessoal de N\'{\i}vel Superior
 - Brazil (CAPES) - Finance Code 001. C.A. also acknowledges the partial financial support of 
 Conselho Nacional de Desenvolvimento Cient\'{\i}fico e Tecnol\'ogico (CNPq), Brazil (grant 308347/2025-0 and  Universal 406820/2025-2), 
and Funda\c c\~ao de Amparo \`a Pesquisa do Estado do Rio de Janeiro (FAPERJ), Brazil (CNE E-26/204.130/2024). L.M. also acknowledges support from a Simons Foundation targeted grant to Instituto Balseiro, Argentina.
E.C. acknowledges the financial support provided by the Center of Advanced Systems Understanding (CASUS), which is financed by Germany’s Federal Ministry of Education and Research (BMBF) and by the Saxon Ministry for Science, Culture and Tourism (SMWK) with tax funds on the basis of the budget approved by the Saxon State Parliament.  E.H-G. acknowledges financial support by the Spanish Ministerio de Ciencia, Innovaci\'on y Universidades
(MICIU/AEI/10.13039/501100011033) through the Maria de Maeztu project CEX2021-001164-M.

\bibliographystyle{IEEEtran}
\bibliography{ref} 

\onecolumngrid

\pagebreak

\newpage
\setcounter{figure}{0}
\renewcommand\thefigure{S\arabic{figure}}    

\setcounter{equation}{0}
\renewcommand\theequation{S\arabic{equation}}    

\setcounter{section}{0}
\renewcommand\thesection{S\arabic{section}}  

\begin{center}
    \textbf{\Large  Supplementary material for "Morphological..."}\\
    E. H. Colombo, L. Menon,  E. Hernandez-Garcia, C. Anteneodo
\end{center}

\section{Supplementary text}

\subsection{Population persistence}

 The linearized population dynamics defined in Eq.~(\ref{eq:main}) can be exactly solved.  Following standard approaches, we first decomposed the dynamics into spatial and  temporal components, $\rho(\mathbf{r},t)=S(\mathbf{r})T(t)$, such that
\begin{align} 
dT/dt &=  \lambda T,\\ \label{eq:sep}
D\nabla^2 S  + aS \equiv {\mathcal{L}}S &=   \lambda S\,, 
\end{align}
where the dynamics is subjected to the Dirichlet boundary condition
 $\rho|_{x\in \mathcal{B}}=0$ on the domain boundary $\mathcal{B}$  and $\lambda$ is an eigenvalue that determines the population growth.
The general solution can then be written as an eigenfunction expansion,
\begin{equation}
\rho(\mathbf{r},t) 
= \sum_{i}^{\infty} 
S_i(\mathbf{r}) e^{ \lambda_i t}  \, ,
\end{equation}
where we assume $\lambda_i>\lambda_{i+1}$.  
Extinction risk can then be characterized by the maximum eigenvalue 
$\Lambda \equiv \max{\{\lambda_i\}} =  \lambda_0 $.

Exact calculations for circular and  rectangular patches have been worked out, providing first insights on how eigenvalues change as a function of the area and boundary of the patch. For these two shapes,  the extinction threshold occurs for a critical area $g_c$, which in dimensional units translate to a critical patch area $A_c = g_c~D/a$. The geometric factor $g_c$ increases as we deviate from circularity~\cite{holmes1994partial}, as we detail in the following: \\

\noindent
{\bf Circular patch.} For example, for a circular patch of radius $R$,  
the fate of the population is governed by the largest growth rate, which corresponds to 
\begin{align} \label{eq:circular}
 \Lambda = a-D(\alpha_{01}/R)^2,
\end{align} 
where $\alpha_{01}\simeq 2.40483$ is
the first zero of the 0th-order Bessel function. 
Hence, the population survives if $\Lambda\ge 0$, that is if $R\ge \alpha_{01}\sqrt{D/a}$, or, in terms of the area, when $A \ge A_c = g_c D/a$, with 
\begin{align} 
\label{eq:circle}
    g_c\equiv \alpha_{01}^2\pi .
\end{align}

\noindent
{\bf Rectangular patch.} In a rectangular patch of sides $L_x$ and $L_y$, with 
$L_y/\gamma=L_x=L$, we have
\begin{equation}     
\Lambda= a-D(1+1/\gamma^2)(\pi/L)^2, 
\label{eq:rectangular}
\end{equation}
and the population survives if the area $A =\gamma L^2 \ge A_c=g_c D/a$, with 
\begin{equation} 
\label{eq:rectangle}
 g_c = (\gamma+1/\gamma)\pi^2.
\end{equation} 
Therefore, note that the rectangular shape is optimal  to support a population (in the sense that population persists for a smaller area) for $\gamma=1$, that is, the square of side $L$,  for which $ g_c = 2\pi^2$.    

Comparing Eqs.~(\ref{eq:circle}) and (\ref{eq:rectangle}), it is clear that the circular shape produces the smallest critical area---a well-established mathematical result~\cite{bookcantrell2004spatial}. 
In fact, it is a proven mathematical result that for the considered dynamics,  deviations from the circular shape imply an increase in the dominant eigenvalue, which imposes an increase in the critical patch area needed for survival.
In general, the exact values of $\Lambda$ can only be worked numerically~\cite{cantrell2001predator}.

\subsection{Continuum-limit extrapolation method}

Following an established approach~\cite{eingengrombacher2015} to solve the eigenvalue problem (\ref{eq:sep}), we begin by discretizing the continuous reaction–diffusion system in Eq.~(\ref{eq:main}) in space and time. The scalar field $\rho(x,y)$  is restricted to  a regular square lattice with $N \times N$ grid points and spacing $\Delta x$  in both directions. 
 The values  of the field at the  discrete  coordinates $(x_i, y_j)$, with $i,j = 1, \dots, N$, are denoted by $\rho_{i,j}$.
In order to express the dynamics in standard matrix form, the field values are arranged column-wise through the vectorization,  
\begin{equation}
\tilde\rho = (\rho_{1,1}, \rho_{2,1}, \dots, \rho_{N,1}, \rho_{1,2}, \rho_{2,2}, \dots, \rho_{N,2}, \dots, \rho_{N,N})^T 
\end{equation}

Under this representation, the discrete Laplacian becomes an 
$N^2 \times N^2$ block–tridiagonal matrix,
\begin{equation}
\label{eq:LaplacianMatrix}
\widetilde{\nabla}^2 =
\frac{1}{\Delta x^2}
\begin{bmatrix}
T      & I_N   & 0     & \cdots & 0 \\
I_N    & T     & I_N   & \cdots & 0 \\
0      & I_N   & T     & \ddots & \vdots \\
\vdots & \vdots& \ddots& \ddots & I_N \\
0      & 0     & \cdots& I_N    & T
\end{bmatrix},
\end{equation}
where $I_N$ is the $(N\times N)$ identity matrix  and 
$T$ is the $N\times N$ tridiagonal matrix 
\begin{equation}
T =
\begin{bmatrix}
-2 & 1 & 0 & \cdots & 0 \\
1 & -2 & 1 & \cdots & 0 \\
0 & 1 & -2 & \ddots & \vdots \\
\vdots & \vdots & \ddots & \ddots & 1 \\
0 & 0 & \cdots & 1 & -2
\end{bmatrix}.
\end{equation}

To incorporate the Dirichlet boundary conditions to an arbitrary spatial domain $\Omega$, 
we define a binary mask $m_{ij}$, namely
\begin{equation}
m_{ij} =
\begin{cases}
1, & \text{if } (x_i, y_j) \in \Omega, \\[4pt]
0, & \text{otherwise,}
\end{cases}
\end{equation}
and the associated  $N^2 \times N^2$ diagonal matrix $D_m = \mathrm{diag}(m_{ij})$, leading to a modified operator,
\begin{equation}
\widetilde{\nabla}^2 = D_m \widetilde{\nabla}^2 D_m.
\end{equation}

Eq.~(\ref{eq:main}) can then be written as 
\begin{align} \notag
   \frac{d}{dt} \tilde\rho(t) &=  (a+D\widetilde{\nabla}^2)\tilde\rho \\
   &= \tilde{\mathcal{L}}\tilde\rho\, ,
\end{align}
and,
decomposing the discrete operator $\widetilde{\mathcal{L}}$ into its eigenvectors, $v_i$,  and  eigenvalues, $\lambda_i$, the general solution extracted:
\begin{align}
 \tilde\rho(\mathbf{r},t) &= \sum_{i}^{\infty} c_iv_i e^{ \lambda_i t}  \, ,
\end{align}
where $c_i$ are coefficients that set the initial conditions.
Important to our analysis is the set of eigenvalues $\{\lambda_i\}_{i<N}$, in particular, the dominant eigenvalue $\Lambda(\Gamma|N) = \max(\{\lambda_i\}_{i<N})$,  for a given patch morphological configuration, $\Gamma$, captured at resolution $N$. For that, we used the implicitly restarted Lanczos method~\cite{lehoucq1998arpack} implemented in the \textit{SciPy} Python library~\cite{virtanen2020scipy} to extract the eigenvalues and select for the largest one.

Noting that the computational cost to compute the dominant eigenvalue in this sparse-matrix problem, measured as runtime, scales as $\mathcal{O}(N^2)$ with resolution, we identify that gains in accuracy come at a steep computational expense. Therefore, to circumvent this issue, we combine results for $\Lambda$  for  the maximum but still feasible resolution, $N_{\textrm{max}}=2000$,  with  additional computationally-cheap values extracted at lower resolutions, e.g. $N \in \{100,200,500,1000,2000\}$.

\begin{figure}[h!]
    \centering
\includegraphics[width=0.5\textwidth]{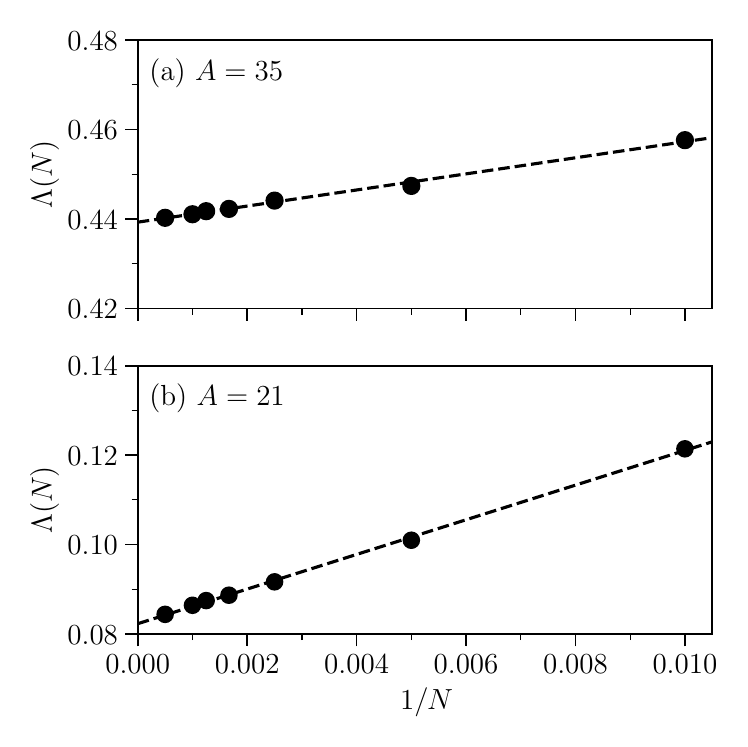}
    \caption{ Eigenvalue  $\Lambda(N)$ plotted as a function of $1/N$, where $N$ is the linear number of grid points,  for  patch boundaries  specified by Eq.~(\ref{eq:fractal_boundary}). 
    The area parameter for 
     each boundary is shown in the legend of each figure. 
    Dashed lines represent the corresponding best linear fits.  
   The asymptotic values, 
   corresponding to the vertical axis intercept, are $\Lambda(\infty) \simeq 0.439$, 
   for (a) $A=35$  and $\Lambda(\infty) \simeq 0.082$, for (b) $A=21$.  
   In both cases, the rugosity is given by $\delta=0.6$. We used $a=D=1$. }
\label{fig:extrapolacao}
\end{figure}

We, then, perform a  Richardson extrapolation \cite{richardson1911,richardson1927} that leverages on low resolution results to go beyond $N_\textrm{max}$. Specifically, the extrapolation incorporates the fact that for  sufficiently high resolution  
$\Lambda$ should converge asymptotically to limiting value, $\Lambda(\Gamma|\infty)$,  following the scaling $\Lambda (\Gamma|N) = \alpha/N + \Lambda(\Gamma|\infty)$ \cite{richardson1911,richardson1927}. This is valid for any boundary contour, even fractal-like (with finite resolution) ones, as the spectrum would have compact support when $N$ is taken large enough. In the case of our fractal-like model, $N$ should be sufficiently large to resolve the boundary oscillations of the $n_M$ modes combined. As shown in Fig.~\ref{fig:extrapolacao}, the convergence is well behaved, as required in the Richardson extrapolation scheme. The scaling law is fitted to our synthetic data, and the best values of $\alpha$ and $\Lambda(\Gamma|\infty)$ are extracted. The extrapolated values, which are used in all our results, show a small  but noticeable  quantitative correction relative to $N_\textrm{max}$ of 5--10\%. Therefore, using results for fixed $N$, even  large  ones, might become substantially inaccurate quantitatively, although results should remain the same qualitatively.

\newpage
 \section{Supplementary figures}

\end{document}